\documentstyle[12pt]{article}
\def\eqbegin         {  \begin{eqnarray}  }
\def\eqend           {  \end{eqnarray}  }
\def\sectionnumbering {\setcounter{equation}{0}\renewcommand{\theequation}{\arabic{section}.\arabic{equation}}}
\def\lamda        { \lambda }
\def\iPhi         { {\mit \Phi}}  
  
\def\iDelta       { {\mit \Delta}}
\def\iLamda       { {\mit \Lambda}}
\def\del          { \partial }

\title{Multiple Edge Partition Functions For 
Fractional Quantum Hall States}
\vspace{50mm}
\author{Kazusumi Ino\thanks{e-mail:ino@kodama.issp.u-tokyo.ac.jp}}
\date{}

\begin{document}
\pagestyle{myheadings}
\markright{}
\thispagestyle{empty}
\thispagestyle{empty}
\begin{center}
\vskip .6 in
{\LARGE Multiple Edge Partition Functions for Fractional Quantum Hall States} \\
\vskip 0.2in
Kazusumi Ino 
\vskip 0.1in

{\it Institute for Solid State Physics, University of Tokyo} \\
{\em ,Roppongi
 7-22-1,  Minatoku,  Tokyo,  106, Japan} \\
\end{center}

\vspace{10mm}

\begin{abstract}
We consider the multiple edge states of the Laughlin state 
and the Pfaffian state. These edge states are globally 
constrained through the operator algebra of 
conformal field theory in the bulk.  We analyze  
these constraints by introducing an expression of
 quantum hall state by the chiral vertex operators 
and obtain the multiple edge partition functions 
 by using the Verlinde formula. 
\end{abstract} 
\vskip 2.0in
\begin {center}
{\bf To appear in Int.J.Mod.Phys.A    }
\end{center}
\newpage
\pagenumbering{arabic}
\section{Introduction}
\sectionnumbering
Fractional quantum hall states \cite{Tsui}
\cite{Lagh} have edge excitations \cite{Stone}\cite{Wen}
\cite{AC1} as their low energy excitations. 
They are the infinitesimal distortion of the quantum hall droplet.
The partition functions for these 
edge excitations have been discussed for the disk and 
the  annulus geometry \cite{Wen}\cite{AC1}\cite{WenWu2}\cite{Milo}\cite{AC2}.
It is natural to generalize the discussions to   edge excitations of  
 quantum hall state on a region with multiple boundaries. 
 In this paper,  we would like to  calculate the multiple edge 
 partition function for these edge excitations, especially  
 for the Laughlin state and the Pfaffian state. 
The wavefunctions for these quantum hall states are known to be 
correlation functions of extending fields of 
certain rational conformal field theories \cite{MorRe}\cite{Cris}\cite{Fubi}
\cite{Flo}. 
Each edge excitation is generated on the primary field on each edge, 
which is determined by the bulk in the case of disk, but has degrees of
freedom in general. These degrees of freedom  are
 globally constrained by the chiral operator algebra
 of the bulk conformal field theory. In view of  this, 
we introduce an expression of quantum 
 hall states in terms of the chiral vertex operators \cite{MorSei} 
  to deal bulk and edge states simulataneously.  
By using this method, we find a relation between
 the partition functions for $n_b$ and $(n_b-1)$ boundaries, 
which is diagonalized by using the Verlinde formula \cite{Ver}. 
We obtain the partition functions by using this relation. 

The organization of the paper is as follows. 
In Sec. 2, we first review the relation between the 
 bulk and the edge state in the Laughlin state on a disk. 
We introduce an  expression in terms of the chiral vertex operators and extend 
it to the general cases with multiple boundaries. We get a relation between
 the partition functions for $n_b$ and $(n_b-1)$ boundaries,  
 which is diagonalized by the Verlinde formula of rational torus. 
The explicit form of the multiple edge partition function 
for the Laughlin state in terms of the matrix elements of modular 
transformation is obtained.
In Sec. 3, we extend the method introduced in Sec. 2 to the Pfaffian state.
We again get a relation between 
 the partition functions for $n_b$ and $(n_b-1)$ boundaries.
We obtain the multiple edge partition functions by diagonalizing 
 the relation  by the Verlinde formula of the Ising model. 
Sec. 4 discusses generalization to other quantum hall states.  

\section{Edge and Bulk of the Laughlin states }
\sectionnumbering 

Before considering edge states from many-body wavefunctions,  
let us recall the relation of 2-dimesional bulk conformal 
field theory and 1+1 dimensional edge conformal field theory  
 to 2+1 dimensional Chern-Simons theory \cite{Witten}\cite{MorSei2}.   
 Chern-Simons theory on a disk is equivalent to 1+1 
dimensional CFT on its edge. 
This 1+1 dimensional CFT describes the edge excitations of FQH state. 
On the other hand, the Hilbert space of Chern-Simons
 theory  with Wilson lines on a Riemann surface $\Sigma$
 is  the space of conformal blocks of 2 dimensional CFT on $\Sigma$. 
 Wilson lines correspond to primay fields of 2 dimensional CFT.
Then bulk wavefunctions are 
 states in the physical Hilbert space of Chern-Simons theory 
 with Wilson lines.
The bridge between these two CFT is discussed  
in \cite{MorSei2}.
 There, by considering  the effect of shrinking a boundary, 
 it is shown that it eventually 
becomes equivalent to a Wilson line. 
This last fact relates the edge states 
to fields of the bulk conformal field theory. For example,   
edge excitations of a quantum hall state on a disk 
are generated on the primary field at the infinity.   

\subsection{Laughlin states on a disk}
Let us recall the relation between many-body  wavefunctions
and edge excitations in the $\nu=\frac{1}{q}$ Laughlin  state on a disk 
($q:$ odd) \cite{Stone}\cite{Wen}\cite{MorRe}\cite{WenWu2}.  
Its ground state is described by the  Laughlin wavefunction 
\eqbegin
\widetilde{\Phi}(z_1,\cdots,z_N)=\prod_{i<j}(z_i-z_j)^{q} {\rm
  exp}\left[ -\frac{1}{4}\sum|z_i|^{2}\right].
\label{laughlin}
\eqend
This wavefunction can be written in terms of a correlator of 
the rational torus with $q$ primary fields. 
It is described by the  chiral boson  field $\varphi$  
compactified on a circle with a rational value of the square of the radius.
It has $q$ primary fields $[\phi_p]=[e^{i\frac{p}{\sqrt{q}}\varphi}]$, 
with U(1) charge $p/q$, 
 $ p \in Z ({\rm mod}\hspace{1mm}  q)$.  The fermionic extending operator
 $\psi_e=e^{i\sqrt{q}\varphi}$ 
has a conformal weight $q/2$ and a unit U(1) charge. 
The Laughlin wavefunction (\ref{laughlin}) can be written 
by this operator as   
\eqbegin 
\widetilde{\Phi}&=& \lim_{z_{\infty}\rightarrow \infty} 
z_{\infty}^{2h_{\rm edge}}\langle \Psi_{\rm edge}(z_{\infty})  \prod_{i=1}^{N}\psi_e(z_i)
{\rm exp}\int \frac{d^2w}{2\pi i}\sqrt{\nu}\varphi(w)  \rangle \nonumber \\
&=& \langle \Psi_{\rm edge}^{\vee}| 
\prod_{i=1}^{N}\psi_e(z_i){\rm exp}
\int \frac{d^2w}{2\pi i}\sqrt{\nu}\varphi(w) |0 \rangle, 
\label{laughlin2}
\eqend  
where the factor with the integrand is the  neutralizing background 
field (or background magnetic field) and will be omitted hereafter,  and   
\eqbegin 
 \Psi_{\rm edge} &=&\phi_{-N}= e^{-iN\sqrt{q}\varphi}, \\
 \langle \Psi_{\rm edge}^{\vee}| &=&  \lim_{z_{\infty}\rightarrow \infty} \langle 0|
 \Psi_{\rm edge}(z_{\infty}) z_{\infty}^{2h_{\rm edge}},  \label{Boundary} 
 \\ 
 h_{\rm edge}&=&\frac{1}{2}N^2q.  
\eqend
Note that (\ref{Boundary}) is the standard definition of the "out" state for 
$\Psi_{\rm edge}^{\vee} $  
in conformal field theory ($\vee$ denotes the conjugate of field ).

Edge excitations are the state of zero-energy for the model Hamiltonian  
(Haldane's pseudopotential) \cite{Hal}, 
\eqbegin 
V=\sum_{l=0}^{q-1}V_l\sum_{i<j}\delta^{(l)}(z_i-z_j).
\label{haldane}
\eqend 
The Laughlin wavefunction (\ref{laughlin}) 
is the exact ground state of $V$ and it is proved that 
all the zero energy states are obtained by acting the symmetric polynomials of 
$z_i$ on the ground-state wavefunction \cite{Hal}.    
In conformal field theory, these edge excitations are described by 
descendant fields of $\Psi_{\rm edge}$ 
generated by the $U(1)$ Kac-Moody algebra $j(z)=\sum_{-\infty}^{\infty}j_n(z-z_{\infty})^{-n-1}$ \cite{WenWu2}, 
\eqbegin 
\Psi_{\rm edge}^{(n_1,n_2,\cdots)} &=&
(j_{-n_1}j_{-n_2}\cdots)\Psi_{\rm edge}, \\
\label{descendant}
 [ j_n,j_m ]&=& n \delta_{m+n} .
\label{U(1)KacMoody}
\eqend
We get edge-excited wavefunctions by inserting 
$\Psi_{\rm edge}^{(n_1,n_2,\cdots)}$ as 
\eqbegin
\widetilde{\Phi}^{(n_1,n_2,\cdots)}&=& \lim_{z_{\infty}
\rightarrow \infty} z_{\infty}^{2h_{\rm edge}+2l}\langle
\Psi_{\rm edge}^{(n_1,n_2,\cdots)}(z_{\infty}) 
\prod_{i=1}^{N}\psi_e(z_i)\rangle , 
\label{edge}
\eqend
where $l=\sum n_k$ is the level of the descendant
 $\Psi_{\rm edge}^{(n_1,n_2,\cdots)}$.
This construction generates the space of symmetric polynomials in $z_i's$ i.e. 
all the edge excitations when we take 
the thermodynamic limit $N \rightarrow \infty$ \cite{Stone}
\cite{Wen}\cite{WenWu2}.

Although edge excitations generate all the zero-energy 
states of $V$, we can add the term proportional to the 
total angular momentum $L$ of electrons.
This is a natural assumption for the confining potential,  since   
$L$ is nothing but the kinetic energy of edge excitatons.     
We can compute the eigenvalue of $L$ on $\widetilde{\Phi}$
 by transforming $z \rightarrow \lamda z$,
$\lambda=e^{i\theta}$. 
In (\ref{laughlin2}),  as $z \rightarrow \lamda z$, we get 
$\Psi_{\rm edge}\rightarrow \lamda^{-h_{\rm edge}}\Psi_{\rm edge}$ 
 and $\psi_e \rightarrow 
 \lamda^{-h_e}\psi_e$, $\widetilde{\Phi}(\lamda z)= 
\lamda^{M_0}\widetilde{\Phi}$  with 
$M_0=h_{\rm edge}-Nh_e=\frac{1}{2}qN(N-1)$. 
 Thus the total angular momentum of 
 the state  $\widetilde{\Phi}$ is $M_0=\frac{1}{2}qN(N-1)$. 
 Likewise in (\ref{edge}), 
$\Psi_{\rm edge}^{(n_1,n_2,\cdots)} \rightarrow \lamda^{-h_{\rm
    edge}-l}\Psi_{\rm edge}^{(n_1,n_2,\cdots)}$,   
we get $\widetilde{\Phi}^{(n_1,n_2,\cdots)}(\lamda z)=\lamda^{M_0+l}
\widetilde{\Phi}^{(n_1,n_2,\cdots)}$ .  Thus the total angular momentum of 
$\widetilde{\Phi}^{(n_1,n_2,\cdots)}$ is  $M_0+l$.  
The number of state at each eigenstate of total
angular momentum is given by the partition number of $l$. 
Therefore, the partition 
function of edge excitations  on disk at the inverse temperature 
$2\pi\tau=i\beta$ is given by 
\eqbegin 
Z^{\rm disk}(\tau)={\rm Tr}\left( e^{2\pi i\tau(L-\frac{c}{24})}\right)=
\frac{\omega^{M_0}}{\eta(\tau)} ,\\ 
\eta(\tau) = \omega^{\frac{1}{24}}\prod_{n=1}^{\infty}(1-\omega^{n}),
\hspace{4mm}  \omega={\rm exp}(2\pi i\tau), 
\label{partition}
\eqend   
where $c=1$ and $-\frac{c}{24}$ is a Casimir energy.  
This partition function is valid in the thermodynamic limit
 $N \rightarrow \infty$.    
   
Now let us see these aspects of the quantum hall state 
$\widetilde{\Phi}$ of $N$ electrons 
 from the constructive point of view.  Suppose that  we make a disk-like 
 subregion $D_1 $ which contains only one electron. Then from (\ref{laughlin2})
  the state at the boundary $\del D_1$ of $D_1$  is 
$\Psi_{\rm edge}^{\vee}=\phi_{q}$.
Next, we make an annulus-like region $D_2$ around $D_1$  which again  
contains one electron. Then the state of the outer boundary of $D_2$ (i.e. $
\del(D_1\cup D_2)$)  is $\Psi_{\rm edge}^{\vee}=\phi_{2q}$. Similarily 
we continue take annulus-like regions $D_3 \cdots$
 to end up with the $N$-th region $D_N$.  The  state at the boundary  
   $\del(\bigcup_{i=1}^{K}D_i)$ is 
$\Psi_{\rm edge}^{\vee}=\phi_{Kq}$.
 The essential ingredient which this construction requires  is 
{\it fusion rules} of conformal field theory. 
The fusion rules of the rational torus are 
\eqbegin 
\phi_r \times \phi_s = \phi_{r+s}. 
\label{u1fusion}
\eqend 
The  procedure above is summarized as follows  :
(i) Take a quantum hall droplet with the edge state $\Psi_{\rm edge}^{\vee}$. 
(ii) Enlarge this droplet by  surrounding it with an annulus-like 
quantum hall liquid with one field $\phi$ .
(iii) Then the new edge state is $\Psi_{\rm edge,new}^{\vee}$ determined  
by the fusion rule   
\eqbegin 
\phi \times \Psi_{\rm edge}^{\vee}=\Psi_{\rm edge, new}^{\vee}.
\eqend 
There is a concept which represents this constrution explicitly.
It is the "chiral vertex operator"  $\iPhi^{i}_{j k}(z) :[\phi_i] 
\rightarrow {\rm Hom([\phi_k]\ \rightarrow [\phi_j])}$ \footnote{
see \cite{MorSei} for detail. }.
 It represents the three holed sphere with 
the fields $\phi_i, \phi^{\vee}_j, \phi_k$ on each hole. 
In our context,  $\iPhi^{i}_{j k}(z)$ represents the annulus with 
the field $\phi_i$ inserted at $z$ with the inner and outer edge states 
$\phi_k$ and   $\phi_j^{\vee}$ respectively . It implies 
$\phi_i \times \phi_k=\phi_j$.  
Then the quantum hall state $\widetilde{\Phi}$  is 
expressed according to 
 the above construction as \footnote{Preliminary description  
 of quantum hall states in terms of  chiral vertex operators is  
discussed in \cite{Ino}.  }  
\eqbegin 
\widetilde{\Phi}=\iPhi^{q}_{Nq,(N-1)q}(z_1)\cdots
\iPhi^{q}_{2q,q}(z_{N-1})\iPhi^{q}_{q,0}(z_N).
\eqend 
For the quantum hall state with edge excitations 
$\widetilde{\Phi}^{(n_1,n_2,\cdots)}$,  we have 
\eqbegin 
\widetilde{\Phi}^{(n_1,n_2,\cdots)}&=&\iPhi^{q}_{b,(N-1)q}(z_1)\cdots
\iPhi^{q}_{2q,q}(z_{N-1})\iPhi^{q}_{q,0}(z_N), \\ 
b^{\vee}&=& (j_{-n_1}j_{-n_2}\cdots)e^{-iN\sqrt{q}\varphi}.
\eqend
We define an operator $M$ on chiral vertex operators by 
\eqbegin 
M\iPhi_{jk}^{i}=(-\iDelta_i+\iDelta_j-\iDelta_k)\iPhi_{jk}^{i} ,   
\eqend 
where $\iDelta_m$ is the conformal weight of field $\phi_m$. 
We demand that it satisfies the Leibnitz rule when it acts
 on the product of chiral vertex operators. Then 
\eqbegin 
M\widetilde{\Phi}&=&M_0\widetilde{\Phi},  \\
M\widetilde{\Phi}{(n_1,n_2,\cdots)}&=&(M_0+l)\widetilde{\Phi}, 
\eqend
where $M_0=\frac{1}{2}qN(N-1)$,  and $l=\sum_{i}n_i$ is the level of 
the descendant field $b$. Thus,  $M$ coincides with   
the total angular momentum operator $L$ when it acts on the space of 
the monomials of chiral vertex operators which express 
$\widetilde{\Phi}^{(n_1,n_2,\cdots)}$. Let us denote the vector space 
spanned by such monomials  as $\Omega_{\rm edge}$.  The partition 
function (\ref{partition}) can now be written as 
\eqbegin 
Z^{\rm disk}={\rm Tr}_{\Omega_{\rm edge}}
 \left( e^{2\pi i\tau(M-\frac{c}{24})} \right).
\eqend

\subsection{ Laughlin states on an annulus}

We consider a Laughlin state 
$\widetilde\Phi(z_1,\cdots,z_N)$ on an annulus $\widetilde{A}$, 
with $N$ electrons. 
We can adjust the velocities of edge excitations on each edge  
 to treat the energy of  each boundary on  equal footing. 
To divide the quantum hall state into components as 
in the previous section,  we first make   
 an annulus-like region $D_1$ (instead of disk) 
 around the inner boundary of $\widetilde{A}$
  which contains only one electron. 
Then we make a series of annulus-like regions,  $D_2\cdots D_N$,  
each of which contains  one electron respectively. 
 Eventually, we end up with  the 
 following expression for $\widetilde\Phi(z_1,\cdots,z_N)$: 
\eqbegin 
 \widetilde{\Phi} =
\iPhi_{\beta_1,  d_1}^{q}(z_1)\cdots \iPhi_{d_{N-1} \beta_2}^{q}(z_N), 
\eqend 
where $\beta_1$ and $\beta_2$ are the primary fields on the outer  
and the inner edges, respectively. The 
edge excitations are  generated by descendant
fields of them in the thermodynamic limit. 
$d_1 \cdots d_N$ are internal states determined by $\beta_1$ and $\beta_2$ . 
On contrary to the case of disk, we can't specify $\beta_1$ and 
$\beta_2$ uniquely.
However, we see that they must satisfy the following condition from
 the U(1) fusin rules:
\eqbegin 
\beta_1= \beta_2+Nq.
\eqend
From this constraint, the general edge states are given by 
\eqbegin
\beta_1=\lamda+m_1q, \hspace{4mm} \beta_2=\lamda+m_2q  \\
 m_1,m_2\in Z,  m_1-m_2=Nq,    \hspace{3mm} \lamda=0, \cdots, q-1 .
\label{sector}
\eqend 
Let $j^{1}_n$ and $j^{2}_n$ be the generators  of U(1) Kac-
Moody algebra on each boundary respectively. Then,  the quantum hall
state with edge excitations are expressed as    
\eqbegin
\widetilde{\Phi}^{(n_1,n_2, \cdots \overline{n}_1,\overline{n}_2,\cdots)} =
\iPhi_{b_1,  d_1}^{q}(z_1)\cdots \iPhi_{d_{N-1} b_2}^{q}(z_N),\\
b_1^{\vee} = (j^{1}_{-n_1}j^{1}_{-n_2}\cdots)e^{-i\beta_1\phi/\sqrt{q}}, 
\hspace{4mm}
b_2 = (j^{2}_{-\overline{n}_1}j^{2}_{-\overline{n}_2}\cdots)
e^{i\beta_2\phi/\sqrt{q}},
\eqend 
where $
l_1=\sum_i n_i$ and  
$l_2=\sum_i \overline{n}_i $ are the level on each edge.
The action of $M$ on $\widetilde{\Phi}^{(n_1 \cdots
  \overline{n}_1\cdots)} $ becomes  
\eqbegin 
M\widetilde{\Phi}
=\left(\frac{1}{2}qN(N-1)+(m_2q+\lamda)N+
l_1-l_2\right)\widetilde{\Phi}, 
\label{pseudo} 
\eqend
which shows $M$ coincides with the angular momentum also in this case. 
On the other hand, we must treat the two boundaries equally  
to consider the energy of the states.  
To this end, we introduce  another expression  
of $\widetilde{\Phi}$ in terms of the chiral vertex operators. 
One starts with  a disk like region $D_1$ on $\widetilde{A}$
 which contains only one  electron. 
Then we make a series of annulus-like regions $D_2\cdots D_N$,  
each of which contains  one electron respectively. 
To produce the two edges, it is necessary to insert a disk with two holes, $C$ 
in the series, which contains no electrons.  Let us denote the 
  chiral vertex operator  corresponding to $C$ as $\iLamda_{j k}^{i}$. 
 Then,  we arrive at the  following expression for $\widetilde{\Phi}$:
\eqbegin 
\widetilde{\Phi}=\iPhi_{\beta_1, d_1}^{q}
\iPhi_{\beta_2^{\vee} d_2}^{q}\cdots
 \iLamda_{c_1 c_3}^{c_2}\cdots
 \iPhi_{d_N 0}^{q}.
\label{pannulus}
\eqend 
We can further arrange this expression by using the worldsheet 
duality of conformal field theory.  Duality results in a set of 
  " duality  transformations" of chiral vertex operators (see \cite{MorSei} 
 for details).  By using duality transformations, we arrange the 
 expression (\ref{pannulus}) to be  
\eqbegin 
\widetilde{\Phi}_{\Lambda}=\iLamda_{\beta_1, Nq}^{\beta_2}
\iPhi_{Nq, (N-1)q}^{q}\cdots \iPhi_{q, 0}^{q}.
\label{annulus}
\eqend
The  states with edge excitations are now expressed as 
\eqbegin 
\widetilde{\Phi}_{\Lambda}^{(n_1,n_2,\cdots,\overline{n}_1,\overline{n}_2\cdots)}
= \iLamda_{b_1, Nq}^{b_2}
\iPhi_{Nq,(N-1)q}^{q}\cdots \iPhi_{q, 0}^{q}.
\label{annuedge}
\eqend 
We will denote the vector space spanned by 
$\widetilde{\Phi}_{\Lambda}^{(n_1,n_2,\cdots,\overline{n}_1,\overline{n}_2\cdots)}$ as 
$\Omega_{N}$.

We define the action of $M$ on $\iLamda_{jk}^{i}$ by   
\eqbegin 
M\iLamda_{jk}^{i}=(\iDelta_i+\iDelta_j-\iDelta_k)\iLamda_{jk}^{i}, 
\eqend 
and  introduce another operator $\overline{M}$ by 
\eqbegin 
\overline{M}\iPhi_{jk}^{i}&=&(\iDelta_i+\iDelta_j-\iDelta_k)\iPhi_{jk}^{i}, \\
\overline{M}\iLamda_{jk}^{i}&=&(\iDelta_i+\iDelta_j-\iDelta_k)\iLamda_{jk}^{i}.
\eqend 
We define the energy for edge states by $M_E=\frac{1}{2}(M+\overline{M})$ 
to cancel the contribution from the bulk.   
The value of $M_E$ on 
$\widetilde{\Phi}_{\Lambda}^{(n_1,\cdots,\overline{n}_1,\cdots)}$ is 
\eqbegin 
M_E
\widetilde{\Phi}_{\Lambda}^{(n_1,\cdots,\overline{n}_1,\cdots)}
=\left(\frac{(m_1q+\lamda)^2}{2q}+l_1+\frac{(m_2q+\lamda)^2}{2q}+l_2 \right)
\widetilde{\Phi}_{\Lambda}^{(n_1,\cdots,\overline{n}_1,\cdots)}.
\label{pseudo2} 
\eqend
We see that $M_E$ coincides with  the sum of the " pseudoenergy" of
 \cite{Milo}.  
Also, $Q\equiv\frac{1}{q}(\overline{M}-M)$ 
acts as the total charge operator 
\eqbegin 
Q
\widetilde{\Phi}_{\Lambda}^{(n_1,\cdots,\overline{n}_1,\cdots)}
=Nq\widetilde{\Phi}_{\Lambda}^{(n_1,\cdots,\overline{n}_1,\cdots)}.
\label{pseudo3} 
\eqend
which would couple to  the chemical potential.

Now let us consider the annulus partition function 
for $M_E$. For that purpose, 
we follow the discussion of \cite{Milo}.  As the change of 
$\beta$'s  by a $q$ unit is equivalent to adding or removing  
an electron, and the bulk state is not disturbed by this change, 
 we extend a single charge sector to include 
 all the sectors differing by integral charges. It  
 means that we must also consider negative $N$.  
At first sight, this seems to be a contradiction, but it is not.
In conformal field theory, the bulk wavefunction can also 
be reconstructed as the correlation function of $e^{-i\sqrt{q}\varphi}$ 
since there is ambiguity in the sign of charge. 
In other words, the ground state wavefunction for the quantum hall state 
can also be expressed as 
\eqbegin 
\iPhi_{-Nq, -(N-1)q}^{-q}\cdots \iPhi_{-q, 0}^{-q}.
\eqend
When we consider the charge sectors as above, we must take into account 
these expressions too. By defining  $Q\equiv\frac{1}{q}(M-\overline{M})$ 
for negative $N$, we can keep track of the sign of the total charge.  
Thus the partition function we will consider 
is the grand-canonical partition function 
on the space $\Omega_{\rm edge}=\bigoplus_{N=-\infty}^{\infty}\Omega_N$ 
($2\pi\tau=i\beta, 2\pi\zeta=-i\mu\beta$ 
where $\beta=1/k_BT$ is the inverse temperature and 
 $\mu$ is the chemical potential) :     
\eqbegin 
Z^{\rm ann}(\tau,\zeta)={\rm Tr}_{\Omega_{\rm edge}}
\left( e^{2\pi i\tau(M_E-\frac{n_bc}{24})+2\pi i\zeta Q} \right). 
\eqend 
Here the term proportional to the central charge ($n_b=2$) is 
a Casimir energy factor. From (\ref{pseudo2})(\ref{pseudo3}), 
we get  
\eqbegin 
Z^{\rm ann}(\tau, \zeta)=\sum_{\lamda=0}^{q-1} \chi_{\lamda}^{2}(\tau,\zeta)
\eqend
where $\chi_{\lamda/q}$ are the characters of the rational torus
\eqbegin 
\chi_{\lamda/q}(\tau,\zeta)=\frac{1}{\eta}\sum_{m \in Z}
e^{2\pi i\tau\frac{(mq+\lamda)^2}{2q}+2\pi i\zeta(m+\frac{\lamda}{q})}. 
\hspace{4mm} 
\label{chi}
\eqend 
We see that $\chi_{\lamda/q}$ satisfies $\chi_{\lamda/q}=\chi_{-\lamda/q}
=\chi_{(q+\lamda)/q}$.
The contribution of each edge can be distinguished by  
introducing complex conjugate variables. Then the partition function 
becomes 
\eqbegin  
Z^{\rm ann}=\sum_{\lamda=0}^{q-1} \chi_{\lamda/q} 
\overline{\chi}_{\lamda/q},
\eqend 
which is nothing but 
the $\Gamma(2)$ invariant partition function obtained 
in \cite{AC1}\cite{Milo}\cite{AC2}. 

\subsection{Laughlin states with multiple edges}
Let $\widetilde{D}$ be a region which has $n_b$ boundaries $B_1,
 \cdots ,B_{n_b}$, 
and consider a Laughlin state $\widetilde\Phi(z_1,\cdots,z_N)$ 
on $\widetilde{D}$. Let $B_{1}$ be the outer boundary
which encloses $\widetilde{D}$. 
When we  divide the quantum hall state into components as in the previous 
sections, it is necessary to  insert $(n_b-1)$  disks with two holes,  
  $C_1,\cdots, C_{n_b-1}$, all of which contains no electron.
By this procedure,  $D$ is divided into $(N+n_b-1)$ regions $D_1,\cdots,D_N ,
C_1,\cdots, C_{n_b-1}$.  
Then,  $n_b$ regions  among $D_1 \cdots D_N,C_1,\cdots, C_{n_b-1}$ 
have one of $n_b$ boundaries $B_1 \cdots B_{n_b}$ respectively.
As in the previous section, 
each $D_l, l= 1,\cdots N $ corresponds to a chiral vertex operator
 $\iPhi^{i}_{j k}$ with $\phi_i$. Also, we assign  a chiral vertex operator
to $C_m$ and denote them as $\iLamda_{j k}^{i}$,  where $k$ is assigned to  
 the outer boundary of $C_m$.  
As in the case of annulus, we arrange the 
 expression  into the following form 
 by duality transformations:  
\eqbegin 
\widetilde{\Phi}=
\iLamda_{\beta_1 \alpha_1}^{\beta_2}\iLamda_{\alpha_1 \alpha_2}^{\beta_3}\cdots
\iLamda_{\alpha_{n_b-2} Nq}^{\beta_{n_b}} 
\iPhi_{Nq,  (N-1)q}^{q}(z_1)\cdots \iPhi_{q 0}^{q}({z_N}).
\label{medge}
\eqend 
We see that $\widetilde{\Phi}$ is now 
separated into the edge and the bulk parts. 
$\beta_1 \cdots \beta_{n_b}$ are the primary fields on
 which the edge excitations are generated. 
 From the fusion rules of the rational torus , 
 $\beta_1, \cdots, \beta_{n_b} $ are globally constrained by the 
following relation:   
\eqbegin 
\beta_1=\beta_2+\cdots\beta_{n_b}+Nq.
\label{nbsector}
\eqend 
Then the general inequivalent states with edge excitations are  
\eqbegin
\widetilde{\Phi}^{(\{n\})}
&=& \iLamda_{b_1 \alpha_1}^{b_2}\iLamda_{\alpha_1 \alpha_2}^{b_3}\cdots
\iLamda_{\alpha_{n_b-2}, Nq}^{b_{n_b}}
\iPhi_{Nq,  (N-1)q}^{q}(z_1)\cdots \iPhi_{q 0}^{q}({z_N}) \\
b_1^{\vee} &=& (j^{1}_{-n^{1}_1}j^{1}_{-n^{1}_2}\cdots) e^{-i\beta_1 \varphi/\sqrt{q}},\\
b_k &=& (j^{k}_{-n^{k}_1}j^{k}_{-n^{k}_2}\cdots) e^{i\beta_k \varphi/\sqrt{q}}, \hspace{3mm} k=2,\cdots n_b
\eqend
where $(\{n\})=(\{n^{1}_1\cdots\}\{n^{2}_1 \cdots\} \{n^{n_b}_1
\cdots\}) $. Let $\Omega_N$ be the vector space spanned by 
expressions $\widetilde{\Phi}^{(\{n\})}$. 
We take $M_E$ to be the pseudoenergy of the system as the  
generalization from the annulus case. 
To consider the partition function, we gather 
all the charge sectors differing by integer charge as in the case of annulus,  
and consider   the space of the edge states  
$\Omega_{\rm edge}=\bigoplus_{N=-\infty}^{\infty}\Omega_N$.   
From (\ref{nbsector}), the (grand-canonical ) partition function is  
\eqbegin 
Z^{(n_b)}(\tau,\zeta)&=&{\rm Tr}_{\Omega_{\rm edge}}
\left ( e^{2\pi i\tau(M_E-\frac{n_bc}{24})+2\pi i\zeta Q} \right) \\ 
&=& \sum_{\lamda_1-\lamda_2-\cdots-\lamda_{n_b}\equiv 0}
\chi_{\lamda_{1}/q}\cdots \chi_{\lamda_{n_b}/q} 
\label{mlaugh}
\eqend
This partition function can also be obtained from the following method using 
the Verlinde formula \cite{Ver}.
First suppose that $\beta_{n_b}=r \hspace{2mm}({\rm mod} \hspace{1mm} q) ,
\hspace{3mm}r\in\{1,\cdots,q-1\}$.
Then $\alpha_{n_b-2}=r$  ({\rm mod} \hspace{1mm} $q$) and 
$b_1,\cdots b_{n_b-1}$ can be seen as 
$(n_b-1)$ edge states in the presence of $r$ quasiholes. 
Let us denote the space of expressions in terms of  chiral 
vertex operators for these edge states as $\Omega_r^{(n_b)}$,  
and introduce the partition function on $\Omega_r^{(n_b)}$,
\eqbegin 
Z^{(n_b)}_r(\tau,\zeta)&=&{\rm Tr}_{\Omega_{r}^{(n_b)}}
\left ( e^{2\pi i\tau(M_E-\frac{n_bc}{24})+2\pi i\zeta Q} \right)
\eqend
From the discussion above,   $Z^{(n_b)}_r$ is factorized
 by $Z^{(n_b-1)}$ as 
\eqbegin 
Z^{(n_b)}_r=\sum_{r=0}^{q-1} \chi_{s/q}Z^{(n_b-1)}_{r+s}.
\eqend
We see that these relations can be written by use of the fusion rules 
of the rational torus $N_{jk}^{i}=\delta^{(q)}_{j+k,i}$ as  
\eqbegin 
Z_r^{(n_b)}=\sum_{s,t}N^t_{rs}\chi_{s/q} Z_{t}^{(n_b-1)}.
\label{Zeq}
\eqend
To get the explicit formula of the partition function, 
 let us recall some facts about the characters 
 $\chi_{\lamda/q}$ (\ref{chi}) of the rational torus. 
 The modular transformation $S : \tau\rightarrow -\frac{1}{\tau}, \zeta\rightarrow -\frac{\zeta}{\tau} $ acts on 
 $\chi_{\lamda}$ as Fourier transformation:
\eqbegin 
\chi_{\lamda} \rightarrow \widetilde{\chi}_\lamda=\frac{1}{\sqrt{q}}\sum_{\lamda'=0}^{q-1}e^{2\pi i\lamda\lamda'/q} \chi_{\lamda'}. 
\eqend   
The matrix elements of $S$ are therefore 
$S^{k}_{n}=\frac{1}{\sqrt{q}}{\rm exp}(2\pi ikn/q)$.  
As conjectured by Verlinde and proved by Moore and Seiberg,  
 The matrix elements $S_i^{j}$ of modular transformation
   and  the fusion rules $N_{jk}^{i}$ of rational conformal 
 field theory have following 
 relation \cite{Ver}\cite{MorSei}: 
\eqbegin 
N_{jk}^{i}&=& \sum_n S_j^{n} \lamda_k^{(n)} S_n^{\dag i}, \\ 
\label{Ver}
\lamda_k^{(n)}&=&{S_k^{n}}/{S^{n}_{0}}.
\eqend 
Now let us solve the equation (\ref{Zeq}) by using this formula. 
First,  we do the Fourier transformation on $Z_r^{n_b}$,
\eqbegin 
F_r^{(n_b)}&=&\frac{1}{\sqrt{q}}\sum_{\lamda}e^{-2\pi ir\lamda/q}
Z_{\lamda}^{(n_b)}. 
\eqend 
Then,  by using the Verlinde formula (\ref{Ver}), (\ref{Zeq}) is now 
arranged into a  simple formula,  
\eqbegin
F_r^{(n_b)}=\sqrt{q}\widetilde{\chi}_{r/q}F_r^{(n_b-1)}
\eqend
As $Z_r^{1}$ is nothing but $\chi_{r/q}$,   
$F_r^{1}=\frac{1}{\sqrt{q}}\sum_{\lamda}e^{-2\pi ir\lamda/q}\chi_{r/q}$. 
Thus we get  
\eqbegin 
F_r^{(n_b)}=q^{\frac{n_b-2}{2}}(\sum_{\lamda}e^{-2\pi ir\lamda/q}\chi_{\lamda/q})(\widetilde{\chi}_{r})^{n_b-1}.
\eqend 
By using the inverse Fourier transformation, we obtain the 
partition functions $Z_r^{n_b}$ as 
\eqbegin  
Z_r^{(n_b)}&=&q^{\frac{n_b-3}{2}}\sum_{\lamda=0}^{q-1}
e^{-2\pi ir\lamda/q}F_{\lamda}^{(n_b)}\\
&=& \frac{1}{q}\sum_{\lamda=0}^{q-1}
e^{-2\pi ir\lamda/q}\left(
\sum_{\lamda'=0}^{q-1}e^{-2\pi i\lamda\lamda'/q}\chi_{\lamda/q}\right)
\left(\sum_{\lamda'=0}^{q-1}e^{2\pi i\lamda\lamda'/q} 
\chi_{\lamda'/q} \right)^{n_b-1}.
\eqend 
In particular, we get the grand-canonical partition function for the 
edge excitation of the Laughlin state with $n_b$ boundaries as  
\eqbegin  
Z_0^{(n_b)}&=&\frac{1}{q}\sum_{\lamda=0}^{q-1}\left(\sum_{\lamda'=0}^{q-1}e^{-2\pi i\lamda\lamda'/q}\chi_{\lamda'/q}\right)
\left(\sum_{\lamda'=0}^{q-1}e^{2\pi i\lamda\lamda'/q} 
\chi_{\lamda'/q}\right)^{n_b-1}
\eqend 
This is indeed the partition function in (\ref{mlaugh}).

\section{Edge and Bulk of the Pfaffian State}
\sectionnumbering
The method using the Verlinde formula to obtain the multiple edge partition function is applicable to other quantum hall states based on 
rational conformal field theories. As an example, we'd like to calculate 
the multiple edge partition function of the Pfaffian state
\cite{MorRe}. 
The ground-state wavefunction of the Pfaffan state at the filling fraction 
$\nu=\frac{1}{q}$  ($q$ : even) for an even number $N$ of electrons is 
\eqbegin
{\rm Pfaff}(\frac{1}{z_i-z_j})\prod_{i<j}(z_i-z_j)^{q}{\rm exp}\left[ -\frac{1}{4}
\sum_i|z_i|^2\right] .
\label{Pfaff} 
\eqend  
This state can be written in terms of Majorana-Weyl
 fermion $\psi$ as :  
\eqbegin 
\langle \psi(z_1)e^{i\sqrt{q}\varphi(z_1)} \cdots 
\psi(z_N)e^{i\sqrt{q}\varphi(z_N)}  {\rm exp}\int \frac{d^2z}{2\pi i}
\sqrt{\nu}\varphi(z)\rangle.
\eqend
The minimal fusion algebra including $\psi$ is that of the Ising model.
This algebra is the " center algebra" in the sense of \cite{WenWu2}.
The Ising model has three primary fields,  $1,\psi$ and $\sigma$,  
where $\sigma$ is the spin field. 
The fusion rules of the Ising model are 
\eqbegin \psi\times\psi=1 ,\hspace{2mm}      
  \psi\times \sigma =\sigma  , \hspace{2mm}
  \sigma \times \sigma =1 +\psi .
\label{isfus}
\eqend 
The couplings to  the rational torus  are restricted by 
 the requirement of single-valuedness and non-singularity 
 of wavefunctions in the electron coodinates. 
 This requirement is shown to be 
equivalent to an orbifold construction \cite{Milo} and 
the allowed couplings are  
$\{e^{i\frac{r}{\sqrt{q}}} \}$,
$\{\psi e^{i\frac{r}{\sqrt{q}}} \}$,
$\{\sigma e^{i\frac{2r+1}{2\sqrt{q}}} \}$,
$r=0,\cdots,q-1$. 

We can apply the same technique used in the Laughlin state 
to express the quantum hall state. The Pfaffian  state (\ref{Pfaff})
 on a disk is expressed in terms of  chiral vertex operators  as 
\eqbegin 
\iPhi_{1\psi}^{\psi}(z_1)\iPhi_{\psi 1}^{\psi}(z_2)\cdots\iPhi_{1\psi}^{\psi}
(z_{N-1})
\iPhi_{\psi 1}^{\psi}(z_{N}).
\label{Pfeven}
\eqend
where the indices for the rational torus are omitted. 
We see that the contribution 
to the edge state from the Ising model is $1$.  
It is also possible to consider 
the Pfaffian state for the odd number of electrons. In this case, 
we have 
\eqbegin 
\iPhi_{\psi1}^{\psi}(z_1)\iPhi_{1 \psi}^{\psi}(z_2)\cdots\iPhi_{1\psi}^{\psi}
(z_{N-1})
\iPhi_{\psi 1}^{\psi}(z_{N}).
\label{Pfodd}
\eqend
The Ising model contribution to the edge state is $\psi$ in this case. 
The edge excitations 
are generated by the descendant fields of the primary field at each edge
 as in the Laughlin state 
 (for the Ising model, descendant fields are generated by
$\{L_{-n}\},n=1,2,\cdots $ of Virasoro algebra). The operator
$M$ acts as  the angular momentum operator  also in this case.
To give explicit formulas for the partition 
functions , let us recall the Virasoro characters of the Ising model ($\omega=
{\rm exp}(2\pi i\tau)$), 
\eqbegin 
\chi^{\rm MW}_1(\tau)&=&\frac{1}{2}\omega^{-\frac{1}{48}}\left(\prod_{0}^{\infty}(1+\omega^{n+\frac{1}{2}}) +
 \prod_{0}^{\infty}(1-\omega^{n+\frac{1}{2}}) 
 \right), \\
 \chi^{\rm MW}_{\psi}(\tau)&=&
 \frac{1}{2}\omega^{-\frac{1}{48}}\left(\prod_{0}^{\infty}(1+\omega^{n+\frac{1}{2}}) -
 \prod_{0}^{\infty}(1-\omega^{n+\frac{1}{2}}) 
 \right), \\ 
 \chi^{\rm MW}_{\sigma}(\tau)&=&
 \omega^{\frac{1}{24}}\prod_{1}^{\infty}(1+\omega^n). 
\eqend  
Now,  the partition function for a disk is obtained as 
\eqbegin 
Z^{\rm disk}(\tau)=
{\rm Tr}_{\Omega_{\rm edge}}\left(e^{2\pi i\tau(M-\frac{c}{24})} \right)
=\frac{\omega^{M_0}\chi^{\rm MW}_1(\tau)}{\eta(\tau)}
\hspace{4mm}  {\rm for} \hspace{3mm} N \hspace{3mm} {\rm even}, \\
M_0=\frac{1}{2}(qN(N-1)-(N-1)) \\ 
Z^{\rm disk}(\tau)=\frac{\omega^{M_0}\chi^{\rm MW}_{\psi}(\tau)}{\eta(\tau)}\hspace{4mm} {\rm for}\hspace{3mm} N \hspace{3mm} 
{\rm odd}, \\
c=\frac{3}{2}, \hspace{4mm} M_0=\frac{1}{2}(qN(N-1)-(N-1)). 
\eqend 
Next,  let us consider the Pfaffian state on an annulus. Following the
same argument to give (\ref{annulus}),  it is expressed by    
chiral vertex operators as 
\eqbegin 
\iLamda_{\beta_1 \alpha_1}^{\beta_2}
\iPhi_{\alpha_1  \alpha_2}^{\psi}\cdots \iPhi_{\alpha_{N-1}, 1}^{\psi}.
\label{pfann}
\eqend 
In the case of the Laughlin state on an annulus,  we gather all the charge 
sectors differing by integral charges into a single sector. 
For the Pfaffian state, we must gather all the charge sectors differing 
by {\it even} integral charges into a single sector, since electrons 
are paired by the degree of freedom from the Ising model. So,  we  
introduce the following functions for the rational torus: 
\eqbegin 
\chi_{r/q}^{\rm even}(\tau,\zeta)=
\frac{1}{\eta}\sum_{m \in Z_{\rm even}}
e^{2\pi i\tau\frac{(mq+\lamda)^2}{2q}+2\pi i\zeta(m+\frac{\lamda}{q})} \\
\chi_{r/q}^{\rm odd}(\tau,\zeta)= 
\frac{1}{\eta}\sum_{m \in Z_{\rm odd}}
e^{2\pi i\tau\frac{(mq+\lamda)^2}{2q}+2\pi i\zeta(m+\frac{\lamda}{q})} 
\eqend
From these definitions, they satisfy 
\eqbegin 
\chi^{\rm even}_{r/q}=\chi^{\rm even}_{-r/q}=\chi^{\rm even}_{(2q+r)/q}
=\chi^{\rm odd}_{(q+r)/q}, \\
\chi^{\rm odd}_{r/q}=\chi^{\rm odd}_{-r/q}=\chi^{\rm odd}_{(2q+r)/q}
=\chi^{\rm even}_{(q+r)/q}.
\eqend
Also,  we introduce the following functions according to the coupling 
of the Ising model and the rational torus ($a$=even, odd):
\eqbegin 
\chi_{i,r}^{a}&=&\chi_i^{\rm MW}\chi_{r/q}^{a}, \hspace{4mm} i=1,\psi,  \\
\chi_{\sigma,r}^{a}&=&\chi_{\sigma}^{\rm MW}\chi_{(r+1/2)/q}^{a}.
\eqend
For $N$ even, $\alpha_1$ in (\ref{pfann}) is $1e^{\pm iN\sqrt{q}\varphi}$. 
If $\beta_1$ is from the even (odd) sector of the rational torus, 
$\beta_2$ is from  the even (odd) sector and visa versa. 
As in the Laughlin state,  
we define the pseudoenergy by $M_E=\frac{M+\overline{M}}{2}$ 
and the total charge operator by  $Q=\frac{1}{q+1}
(\overline{M}-M)$ for the expressions with positive $N$ and  $Q=\frac{1}{q+1}
(M-\overline{M})$ for the expressions with negative $N$.
    Then,   the grand-canonical partition 
function for $N$ even is  
\eqbegin 
Z^{\rm even}&=& {\rm Tr}_{\Omega^{\rm even}_{\rm edge}}
\left( e^{2\pi i\tau(M_E-\frac{n_bc}{24})+2\pi i\zeta Q} \right)
  \nonumber \\
&=& \sum_{a}\sum_{r=0}^{q-1}\left[(\chi_{1.r}^{a})^2+  (\chi_{\psi.r}^{a})^2+(\chi_{\sigma}^{a})^2 \right],
\eqend 
where $\Omega^{\rm even}_{\rm edge}=\bigoplus_{N:{\rm even}}\Omega_N$.
For $N$ odd, $\alpha_1$ in (\ref{pfann}) is $\psi e^{\pm iN\sqrt{q}\varphi}$. 
If the $\beta_1$ is from the even(odd) sector of the rational torus, 
$\beta_2$ is from  the odd (even ) sector and visa versa. From (\ref{isfus}), 
we obtain 
\eqbegin 
Z^{\rm odd}&=& {\rm Tr}_{\Omega^{\rm odd}_{\rm edge}}
\left( e^{2\pi i\tau(M_E-\frac{n_bc}{24})+2\pi i\zeta Q} \right)
  \nonumber \\
&=&\sum_{r=0}^{q-1} \left[2\chi_{1,r}^{\rm even}\chi_{\psi,r}^{\rm odd}+2\chi_{1,r}^{\rm odd}\chi_{\psi,r}^{\rm even}+2\chi_{\sigma,r}^{\rm even}\chi_{\sigma,r}^{\rm odd} \right],
\eqend
with $\Omega^{\rm odd}_{\rm edge}=\bigoplus_{N:{\rm odd}}\Omega_N$.
By introducing  complex conjugate variables to distinguish two edges, 
we see that the sum of these partition functions 
 $Z^{\rm even}+Z^{\rm odd}$ is  
\eqbegin 
Z^{\rm annulus}=\sum_{r=0}^{q-1}\left[ |\chi_{1,r}^{\rm even}+
\chi_{\psi,r}^{\rm odd}|^2+ 
|\chi_{\psi,r}^{\rm even}+
\chi_{1,r}^{\rm odd}|^2+
|\chi^{\rm even}_{\sigma,r}+\chi^{\rm odd}_{\sigma,r}|^2  \right] \nonumber\\
=\sum_{r=0}^{q-1}\left[ |\chi_1^{\rm MW}\chi_r^{\rm even}+
\chi_{\psi}^{\rm MW}\chi_r^{\rm odd}|^2+ 
|\chi_{\psi}^{\rm MW}\chi_r^{\rm even}+
\chi_1^{\rm MW}\chi_r^{\rm odd}|^2+
|\chi_{\sigma}\chi_{r+1/2}|^2  \right]. 
\eqend
This partition function is the one derived in \cite{Milo} when $\zeta=0$.

Generalization to the case of $n_b$ boundaries is similar to the case
of the Laughlin state. 
The Pfaffian state on a region with $n_b$ boundaries is expressed as 
\eqbegin 
\iLamda_{\beta_1 \alpha_1}^{\beta_2}\iLamda_{\alpha_1 \alpha_2}^{\beta_3}\cdots
\iLamda_{\alpha_{n_b-2} \alpha_{n_b-1}}^{\beta_{n_b}} 
\iPhi_{\alpha_{n_b-1} c_1}^{\psi}\iPhi_{c_1 c_2}^{\psi}
\cdots \iPhi_{c_{N-1} 1}^{\psi}.
\label{medge2}
\eqend 
Let us denote the partition function of edge excitations with
 $\alpha_{n_b-1}=(i,\lamda,a)$ where $i=1,\psi,\sigma$ and $   
 \lamda=0,\cdots, q-1, a={\rm even,odd}$ as $Z_{i,\lamda}^{a,(n_b)}$.
By summing over all the sectors of
 $\beta_{n_b}$, we find the equation satisfied by $Z_{i,\lamda}^{a,(n_b)}$: 
\eqbegin 
 Z_{i,s}^{a_1,(n_b)}=\sum_{jk}\sum_{a_2+a_3=a_1}\sum_{\lamda=0}^{q-1}
 N^{k}_{ij}\chi_{j,\lamda}^{a_2}Z^{a_3,(n_b-1)}_{k,s+\lamda}.
\eqend   
Here $N^{k}_{ij}$ is the fusion rules of the Ising model and $Z^{(n_b-1)}$ is 
the partition function for the remaining $(n_b-1)$ boundaries. 
The Verlinde formula (\ref{Ver}) implies that 
this equation is diagonalized by  the modular transformation. 
The matrix element of modular transformation for the Virasoro 
characters 
$\chi^{\rm MW}_1,\chi^{\rm MW}_{\sigma},\chi^{\rm MW}_{\psi}$ 
of the Ising model are 
\eqbegin 
S^{i}_{j}=\frac{1}{2}\left(
\begin{array}{ccc}
1 & \sqrt{2} &   1\\
\sqrt{2} & 0 & -\sqrt{2}         \\
1 & -\sqrt{2} &  1     \\
\end{array} 
\right).
\eqend
Putting $W_i=\sum_j S^j_i Z_j$, we get 
\eqbegin 
W^{a,(n)}_{i,s}=\sum_{a_1+a_2=a}\sum_{\lamda=0}^{q-1}\left( 
\sum_{j}\lamda_j^{(i)}\chi_{j,\lamda}^{a_1}
\right)W_{i,s+\lamda}^{a_2,(n-1)},  
\label{wlad}
\eqend
where $\lamda^{(i)}_{j}=S^i_j/S^{i}_0.$ 
By introducing
 $\xi_{i,\lamda}^{a}=\sum_{j}\lamda_j^{(i)}\chi_{j,\lamda}^{a}$ i.e. 
\eqbegin 
\xi_{1,\lamda}^{a}&=&\chi_{1,\lamda}^{a}+\chi_{\psi,\lamda}^{a}
+\sqrt{2}\chi_{\sigma,\lamda}^{a}, \\
\xi_{\psi,\lamda}^{a}&=&\chi_{1,\lamda}^{a}+\chi_{\psi,\lamda}^{a}
-\sqrt{2}\chi_{\sigma,\lamda}^{a}, \\
\xi_{\sigma,\lamda}^{a}&=&
\chi_{1,\lamda}^{a}-\chi_{\psi,\lamda}^{a},
\eqend
(\ref{wlad}) becomes 
\eqbegin 
W^{a,(n)}_{i,s}=\sum_{a_1+a_2=a}\sum_{\lamda=0}^{q-1}
\xi_{i,\lamda}^{a_1}W_{i,s+\lamda}^{a_2,(n-1)}.
\eqend
As $W_{i,\lamda}^{a,(1)}=S^{i}_0\xi_{i,\lamda}^{a}$, we get 
:
\eqbegin 
W_{1,s}^{a,(n)}&=&\frac{1}{2}\sum_{p_2,\cdots,p_n}
\sum_{a_{2},\cdots,a_n}
\xi^{a_1}_{1,p_1}\cdots\xi^{a_n}_{1,p_n},  \\ 
W_{\psi,s}^{a,(n)}&=&\frac{1}{2}\sum_{p_2,\cdots,p_n}
\sum_{a_{2},\cdots,a_{n}}
\xi^{a_1}_{\psi,p_1}\cdots\xi^{a_n}_{\psi,p_n}, \\
W_{\sigma,s}^{a,(n)}&=&\frac{\sqrt{2}}{2}\sum_{p_2,\cdots,p_n}
\sum_{a_2,\cdots,a_n}
\xi^{a_1}_{\sigma, p_1}\cdots\xi^{a_n}_{\sigma, p_n}, 
\label{www}
\eqend
with $p_1\equiv s+\sum_{l=2}^{n}p_l \hspace{3mm} ({\rm mod}\hspace{2mm}2q)$ and $ a_1\equiv a+\sum_{l=2}^{n} a_l $.
In terms of these formulas,  the partition functions are obtained as   
\eqbegin 
Z^{a,(n_b)}_{1,s}&=&\frac{1}{2}W_{1,s}^{a,(n_b)}+
\frac{\sqrt{2}}{2}W_{\sigma,s}^{a,(n_b)}
+\frac{1}{2}W_{\psi,s}^{a,(n_b)}, \\
Z^{a,(n_b)}_{\psi,s}&=&\frac{1}{2}W_{1,s}^{a,(n_b)}-
\frac{\sqrt{2}}{2}W_{\sigma,s}^{a,(n_b)}
+\frac{1}{2}W_{\psi,s}^{a,(n_b)}, \\
Z^{a,(n_b)}_{\sigma,s}&=&\frac{\sqrt{2}}{2}W_{1,s}^{a,(n_b)}-
\frac{\sqrt{2}}{2}W_{\psi,s}^{a,(n_b)}.
\eqend 
In particular, we get the partition function for
 the Pfaffian states 
$Z_{1,0}^{\rm even}+Z_{\psi,0}^{\rm odd}$ : 
\eqbegin 
Z_{\rm Pfaff}^{(n_b)}&=&\frac{1}{2}\left(W_{1,0}^{{\rm even},(n_b)}+
W_{1,0}^{{\rm odd},(n_b)}\right) \hspace{1cm}\nonumber \\
&{ }&+ \frac{\sqrt{2}}{2}\left(W_{\sigma,0}^{{\rm even},(n_b)}
-W_{\sigma,0}^{{\rm odd},(n_b)}\right) \nonumber \\ 
&{ }&+ \frac{1}{2}\left(W_{\psi,0}^{{\rm even},(n_b)}+
W_{\psi,0}^{{\rm odd},(n_b)}\right).
\eqend

\section{Discussions}
\sectionnumbering
We obtained the multiple edge (grand-canonical) partition functions 
for the Laughlin state and the Pfaffian state. To deal  
bulk and edge states simultaneously,  we introduced a method 
to express quantum hall states in terms of the chiral vertex operators.  
The constraints result in the relation between the multiple edge 
partition functions, which is diagonalized by the modular transformation. 
These methods are applicable to other states obtained 
from rational conformal field theories. Generally, 
the relation between the (grand-canonical) 
partition functions of edge excitations on 
$n_b$ and $(n_b-1)$ boundaries  is given by  
\eqbegin 
Z_{i}^{(n_b)}&=&\sum_{jk}
 N^{k}_{ij}\chi_{j}Z^{(n_b-1)}_{k},  
\eqend
where $\chi_j$ is the Virasoro character for $\phi_j$  and 
the sum is over the allowed primary fields, which are determined by  
the single-valuedness of wavefunctions in the electron coordinates. 
By using this equation recursively, we have 
\eqbegin  
 Z^{(n_b)}_{i}
 &=&\sum_{j_1,\cdots,j_{n_b}}\sum_{k_1,\cdots,k_{n_b-2}}
  N^{k_{n_b-2}}_{i j_{n_b}}N^{k_{n_b-3}}_{k_{n_b-2} j_{n_b-1}} 
 \cdots N^{k_1}_{k_{n_2} j_3}N^{j_1}_{k_1 j_2}
  \chi_{j_1} \cdots \chi_{j_{n_b}} \\
 &=&\sum_{j_1,\cdots,j_{n_b}} {\rm dim}{\cal H}(i,j_1^{\vee},
 j_2\cdots,j_{n_b}) 
 \chi_{j_1} \cdots \chi_{j_{n_b}},  
\eqend
where  ${\rm dim}{\cal H}(i,j^{\vee}_1,\cdots,j_{n_b})
= N^{k_{n_b-2}}_{i j_{n_b}}N^{k_{n_b-3}}_{k_{n_b-2} j_{n_b-1}} 
 \cdots N^{k_1}_{k_{n_2} j_3}N^{j_1}_{k_1 j_2}$ is the dimension of 
\newline 
${\cal H}(i,j^{\vee}_1,\cdots,j_{n_b})$, 
the space of conformal blocks 
on a sphere with $n_b+1$ marked points with the insertions of 
 fields $ i,j^{\vee}_1,j_2\cdots,j_{n_b}$.  On the other hand, we obtain  
the explicit form  of  $Z_i$ by using  the Verlinde formula (\ref{Ver}) as 
\eqbegin 
Z_{i}^{(n_b)}=\sum_{k}S_i^{k}(\sum_j S_j^{\dag k}\chi_j)\left(\sum_j \frac{S_j^{k}}{S_0^{k}}
\chi_j\right)^{n_b-1}.
\eqend
The actual multiple edge partition function is obtained by combining 
these $Z_i^{(n_b)}$.  The combination depends on the relation of 
bulk and edge states of the given quantum hall state.

It would also be interesting to calculate the multiple edge partition 
functions for the Haldane-Rezayi state\cite{HalRez},
  which is beyond the scope of  rational conformal field theory 
  \cite{MorRe}\cite{WenWu2}\cite{Milo}\cite{GFN}\cite{LeeWen}. 

\subparagraph{Acknowledgement}
The author would like to thank G.R.Zemba, and especially M. 
Flohr for useful suggestions and comments on the manuscript.


\begin{thebibliography}{bl}
\bibitem{Tsui}   D.C.Tsui, H.L.Stormer and A.C.Gossard, Phys.Rev.Lett. {\bf 48}
(1982)1559; For a review, see {\it The Quantum Hall Effect, 2nd.ed.}, 
edited by R.E.Prange and S.M.Girvin(Springer-Verlag,NewYork,1990).
\bibitem{Lagh}     R.B.Laughlin, Phys.Rev.Lett.{\bf 50} (1983)1395.
\bibitem{Stone}  M.Stone, Ann.Phys.(N.Y.) {\bf 207} (1991) 38 ;Phys.Rev.{\bf B 42} (1990)8399.
\bibitem{Wen} for a review see:  X.G.Wen, Int.J.Mod.Phys. {\bf B6} (1992)1711.
\bibitem{AC1}  A.Cappelli, G.V.Dunne, C.A.Trugenberger and G.R.Zemba, Nucl.Phys
{\bf B398}(1993) 531. 
\bibitem{WenWu2}  X.G.Wen, Y.S.Wu,Y.Hatsugai Nucl.Phys. {\bf B422}[FS]
(1994)476; X.G.Wen, Y.S.Wu,  Nucl.Phys. {\bf B419} (1994)455.
\bibitem{Milo} M.Milovanovi$\acute{\rm c}$ and N.Read, Phys.Rev. {\bf B53}(1996)13559.
\bibitem{AC2} A.Cappelli and G.R.Zemba, Nucl.Phys.{\bf B490} (1997) 595.
\bibitem{Hal}    F.D.M.Haldane, Phys.Rev.Lett.{\bf 51} (1983) 645.
\bibitem{MorRe} G.Moore and N.Read, Nucl.Phys. {\bf B360} (1991) 362.    
\bibitem{Cris} C.Cristofano, G.Mauekka, R.Musto, F.Nicodemi, 
Phys.Lett.{\bf B262}(1991)88, 
Mod.Phys.Lett. {\bf A6}(1991)1779, 
Mod.Phys.Lett. {\bf A6}(1991)2217.
\bibitem{Fubi} S.Fubini, Mod.Phys.Lett.{\bf A6}(1991)347.
\bibitem{Flo}  M.Flohr, Mod.Phys.Lett{\bf A11}(1996)55.
\bibitem{MorSei}  G.Moore and N.Seiberg, Phys.Lett.{\bf 212B} (1988) 451
;Nucl.Phys. {\bf B313} (1989) 16
;Commun.Math.Phys. {\bf 123} (1989) 77.
\bibitem{Ver}    E. Verlinde, Nucl.Phys. {\bf B300}[FS22] (1988) 360. 
\bibitem{Witten}   E. Witten, Commun.Math.Phys. {\bf B121}(1989) 351. 
\bibitem{MorSei2}  G.Moore and N.Seiberg, Phys.Lett.{\bf 220B} (1989) 422.
\bibitem{Ino} K.Ino, cond-mat/9702039.
\bibitem{HalRez} F.D.M.Haldane and R.Rezayi,  Phys.Rev.Lett. {\bf 60} (1988) 956. 
\bibitem{GFN} V.Gurarie,  M.Flohr and C.Nayak, Nucl.Phys.{\bf B498} (1997) 513.
\bibitem{LeeWen} J.C. Lee  and X.G. Wen, cond-mat/9705033.

\end{thebibliography}
\end{document}